\documentclass[%
 aip,
 amsmath,amssymb,
 reprint,%
]{revtex4-1}

\usepackage{graphicx}
\usepackage{dcolumn}
\usepackage{bm}

\usepackage[utf8]{inputenc}
\usepackage[T1]{fontenc}
\usepackage{mathptmx}
\usepackage{etoolbox}
\usepackage{float}
\usepackage{xcolor}

\makeatletter
\def\@email#1#2{%
 \endgroup
 \patchcmd{\titleblock@produce}
  {\frontmatter@RRAPformat}
  {\frontmatter@RRAPformat{\produce@RRAP{*#1\href{mailto:#2}{#2}}}\frontmatter@RRAPformat}
  {}{}
}%
\makeatother
\begin{document}

\preprint{AIP/123-QED}

\title[Delta-Doped Diamond via in-situ Plasma-Distance Control]{Delta-Doped Diamond via in-situ Plasma-Distance Control}
\author{Philip Schätzle}
    \email{philip.schaetzle@iaf.fraunhofer.de}

\author{Felix Hoffmann}
\author{Sven Mägdefessel}
\author{Patrik Straňák}
\author{Lutz Kirste}

\author{Peter Knittel}%
\affiliation{ 
Fraunhofer Institute for Applied Solid State Physics IAF, Tullastr. 72, 79108 Freiburg, Germany
}%

\date{\today}

\begin{abstract}
We present an approach to CVD diamond growth, in which the sample is placed at a defined distance from the reactor baseplate, to which the plasma couples. We observe two previously unknown growth regimes. In the first case, the sample is positioned within three to five millimeters of the plasma. This leads to a decreased growth rate, compared to a position inside the plasma and, additionally, to an increased nitrogen incorporation, allowing the fabrication of delta-doped layers with a thickness below 30~nm. In another regime, where the sample is positioned more than 10~mm away from the plasma, no growth is observed. Instead, we assume that a thin layer of nitrogen-rich species on the diamond surface is formed, which is incorporated during the growth of the following layer. This enables the fabrication of delta-doped layers with thicknesses below 10~nm. All doped layers show NV emission, with the intensity correlating with the nitrogen incorporation. The growth techniques could enable the fabrication of highly doped thin films for quantum sensing applications, as well as layers with low NV concentration, for quantum computing. The new approaches are not only applicable to nitrogen incorporation but also to other dopants such as phosphorus, which could open up new avenues for diamond-based electronics.
\end{abstract}

\maketitle

The nitrogen vacancy (NV) center in diamond is a promising candidate for the implementation of quantum technologies, such as quantum sensors and quantum computers~\cite{prawer2008diamond,abobeih2022fault,degen2017quantum, taylor2008high}. The strong photoluminescence (PL) of dense NV layers, as well as the high coherence time of the electron spin, allow for highly sensitive sensing devices~\cite{herbschleb2019ultra}. The NV-center also offers applications in the field of quantum computing, where proximate nuclear spins can be used as additional qubits~\cite{bradley2019ten}. 

These applications place stringent requirements on the diamond host material, including a high structural quality and low concentrations of unintentional defects~\cite{bauch2020decoherence}. One method of fabricating such diamond layers is via chemical vapor deposition (CVD). This technique enables the production of high-quality diamond containing a variety of dopants~\cite{schwander2011review,kalish1999doping}.

One way to grow high-quality diamond via CVD-growth, is through an enclosed sample holder, where the sample surface is at a lower level than the holder surface~\cite{ren2024recent, mokuno2005synthesizing, nad2015growth}. The depth of this enclosed area is typically a few~millimeters~\cite{mokuno2005synthesizing}. Using this technique, the structural quality of thick diamond layers, fabricated at high growth rates, has been improved compared to samples grown on an open holder ~\cite{mokuno2005synthesizing}.

However, some applications such as quantum sensors require thin, so-called delta-doped diamond layers~\cite{Ohno2012, jaffe2020novel}. The sensitivity of such a sensor can be enhanced by a high signal, arising from a high density of NV centers and close proximity to the specimen through the vertical positioning of the centers~\cite{jaffe2020novel}. Moreover, the coherence time could also be improved in lower-dimensional layers, compared to bulk material~\cite{schatzle2024spin}.

Fabricating such thin films is challenging, especially since the growth rate of diamond increases drastically when nitrogen is introduced during CVD-growth~\cite{jin1994effect}. Furthermore, the incorporation efficiency of dopants such as nitrogen and phosphorus in diamond is low~\cite{lobaev2017influence, schatzle2023chemical, samlenski1995incorporation, tallaire2006characterisation}.  Various techniques have been proposed to grow nitrogen-doped diamond at low rates, allowing for a highly controlled layer thicknesses~\cite{Ohno2012, butler2017nanometric, vikharev2016novel, bogdanov2020optical, bogdanov2021investigation, kato2007n}. These include growth at low microwave powers~\cite{Ohno2012} and subsequent electron irradiation~\cite{Ohno2012}, as well as surface termination and subsequent overgrowth~\cite{jaffe2020novel}. Additionally, changes to the diamond reactor itself have been proposed, e.g. a laminar flow reactor, where the gas switching is more abrupt~\cite{bogdanov2020optical}. Based on this, delta-doped layers with a high nitrogen concentration of over 1000~ppm or an NV concentration down to the single-center level have been obtained in high-quality diamond~\cite{jaffe2020novel, Ohno2012}.

In this work, we present an alternative approach for the fabrication of nitrogen delta-doped layers in diamond via chemical vapor deposition. The method involves keeping the diamond sample in an adjustable retention position during growth, which cam improve the quality of the fabricated diamond~\cite{mokuno2005synthesizing}. This method enables control of the growth rate while maintaining a constant plasma composition. Adjustment of layer thickness and nitrogen incorporation is achieved by changing the retention position with respect to the plasma. Using this technique, we can fabricate delta-doped layers that are a few nanometers thick with high nitrogen incorporation. 

\begin{figure*}[t]
\includegraphics[width=0.99\linewidth]{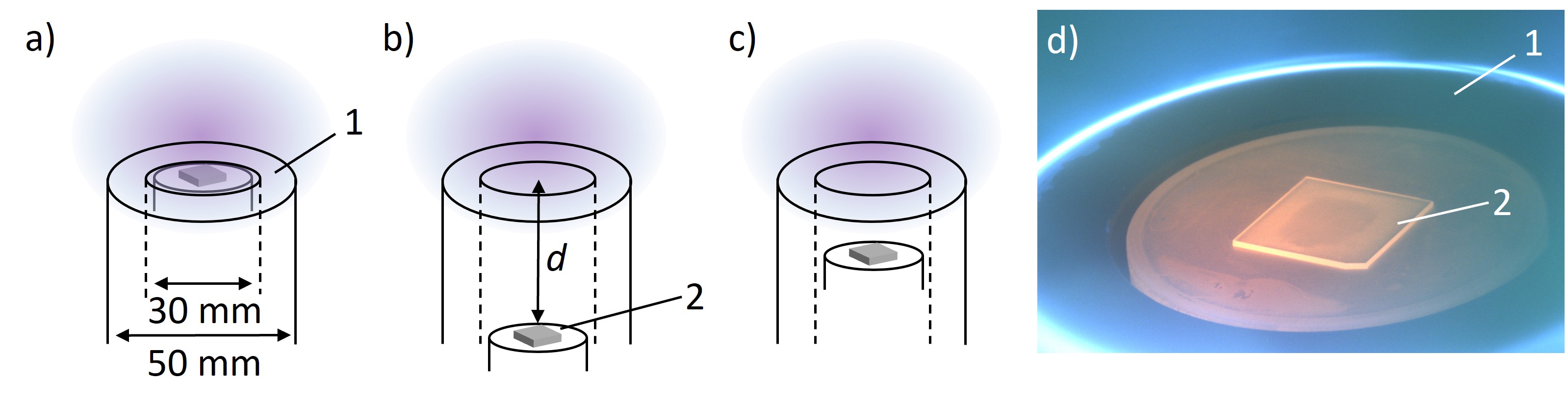}
\caption{Illustration showing the different positions of the sample (2) with respect to the baseplate (1). The sample size is not to scale. a) The sample holder lies at the same level as the baseplate, with the sample in the plasma (reference position). b) The sample holder is far away from the baseplate level, which interrupts the diamond growth process. c) Intermediate positions, where the growth is influenced by the exact position of the sample. d) Image of a sample in the reactor in the reference position, as shown in a).}
\label{fig_reactor}
\end{figure*}

To fabricate said layers, a chemical vapor deposition diamond reactor was used, which is described elsewhere~\cite{schatzle2023chemical}. The main part of this reactor is a substrate lift that enables the sample to be removed and reintroduced to the plasma ball. The term 'plasma ball' refers to the visible portion of the plasma that couples to the molybdenum baseplate. This enables thin film growth with sharp interfaces, since the growth process is interrupted if the distance to the plasma is sufficiently great. To prevent an excessive cooling of the sample, when it is removed from the plasma, a heater that is situated within the sample holder is used. The heater keeps the sample at an elevated temperature; however, the exact temperature of the diamond cannot be determined in the retention position, because there is no line of sight to the attached pyrometer, and the thermocouple attached to the heater is too far from the sample. In this work, we investigate the growth behavior and dopant incorporation in the intermediate regime, where the distance to the plasma ball is small enough to allow plasma species to reach the sample and react, as shown in Figure 1. This position is defined as the distance $d$ from the upper level of the base plate to which the plasma couples. In the following, we refer to the position where the holder is on the same level as the baseplate as the 'reference position'.

Diamond growth was performed using hydrogen and methane as the main source gases. The total gas flow was 800 sccm at a pressure of 130 mbar, resulting in a rapid changeover of the gases in the chamber in less than two minutes after switching. The C/H ratio is 1.75~\% during doped growth. \textsuperscript{15}N was used as a dopant, with an N/C-ratio of 71,000 ppm. Under these conditions, a stack of six doped layers  (samples S1 and S2) and single doped layers (sample S3, S4 and S5) were grown on  (001)-oriented CVD diamond substrates (Element6, electronic grade). The samples were placed on a flat holder with a diameter of 2.54~cm. Using \textsuperscript{12}C-enriched source gas, the grown, doped layers can be distinguished from the substrate below. Between the doped layers and on top of the last doped layer, we fabricated intrinsic layers which have a \textsuperscript{13}C concentration of 1.1~\%, equivalent to the natural abundance. Growth was performed on a first sample (S1), varying the distance to the baseplate, to which the plasma couples, between 20~mm and 0.1 mm during the 5-minute doped growth. The following intrinsic layer was grown for 10~min in the adapted position and subsequently 5~min in the reference position. The growth conditions for one of the doped layers and the subsequent buffer layer are summarized in Table~\ref{tab:table1}. A more detailed description of the growth processes and sample preparation can be found in the Supplementary Material.

\begin{table}
\caption{\label{tab:table1}Growth conditions during the fabrication of the stack of doped layers and intrinsic buffer layers of sample S1.}
\begin{ruledtabular}
\begin{tabular}{cccccc}
Step&CH\textsubscript{4}-conc.&\textsuperscript{13}C-conc.&N/C&position&time\\
&[\%]&[\% of CH\textsubscript{4}]&[ppm]&[mm]&[min]\cr
\hline

1 &1.75 & 0.0 & 71000 & 20 / 10 / 0.1 / &5\\
&&&& 15 / 5 / 3\\
2 &2.0& 1.1 & 0 &20 / 10 / 0.1 /&10\\
&&&& 15 / 5 / 3\\
3 &2.0& 1.1 & 0&0&5\\
\end{tabular}
\end{ruledtabular}
\end{table}

\begin{figure}[t]
\centering
\includegraphics[width=0.99\linewidth]{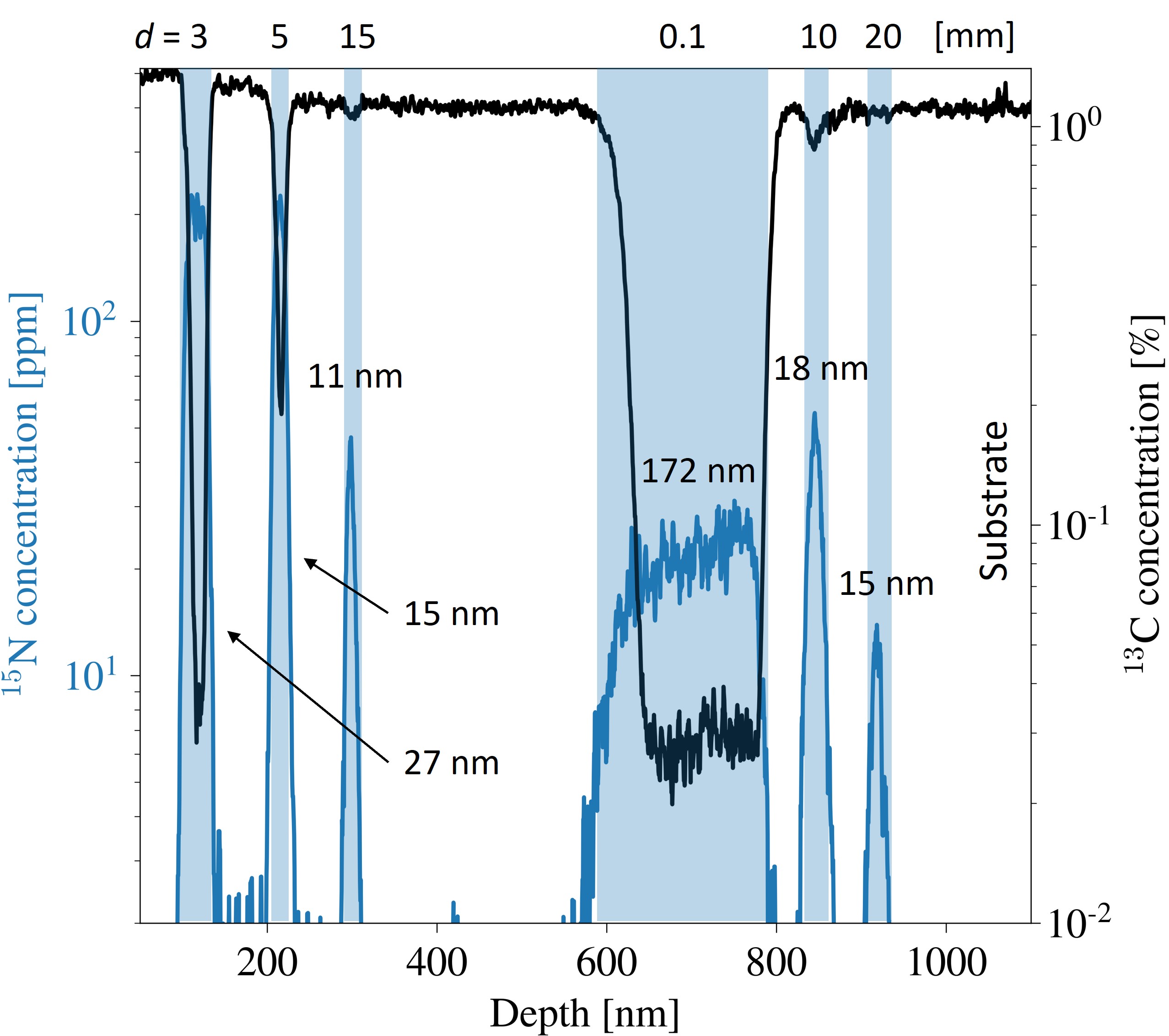}
\caption{ToF-SIMS profile of sample S1, showing the layer stack of six doped layers (blue). The distance $d$ to the base plate was varied in each of the grown layers (20, 10, 0.1, 15, 5, 3~mm), starting from the substrate, as indicated for each layer above the graph. The layers are also indicated by the dip in the \textsuperscript{13}C concentration (black).}
\label{fig_sims_overview}
\end{figure}

To assess the doping of diamond, that was grown in varied growth positions, time-of-flight secondary ion mass spectrometry (ToF-SIMS) measurements were performed. Fig. \ref{fig_sims_overview} shows the layer stack of sample S1.

The measurements revealed significant changes in growth rate and nitrogen incorporation when changing the position of the sample during growth. We distinguish between three different growth modes. The first is the trivial mode, where the sample holder is positioned 0.1 mm below the baseplate level, where the growth conditions are established and well known\cite{schatzle2023chemical}. The second position is slightly below this first position at a distance of 3~mm, where a significant drop in the growth rate and an increase in the nitrogen incorporation were observed. The third regime starts at a distance of more than 15~mm. Here,  nitrogen incorporation is still observed, but no carbon incorporation is indicated by the ToF-SIMS profile, as shown in Fig.~\ref{fig_sims_overview}. It should be noted, that this distance-dependent behavior may be influenced by the reactor geometry, especially by the diameter of the opening in the baseplate.

In a position 0.1~mm below the baseplate level, a layer with a thickness of 172~nm was obtained, as determined from the full width at half maximum of the nitrogen concentration. This corresponds to a growth rate of 2.1~µm/h. The layer is also indicated by the dip in the \textsuperscript{13}C concentration. The temperature during growth was 730~°C, as determined by a pyrometer. On average, the nitrogen concentration in the layer is 22~ppm, corresponding to an incorporation efficiency of $3.1\cdot10^{-4}$, which is in line with literature results~\cite{lobaev2017influence, schatzle2023chemical, samlenski1995incorporation, tallaire2006characterisation}.

In lower positions, that are still close to the plasma (3 to 5~mm), a significant change in growth behavior is observed. The growth rate drops drastically to 320~nm/h, with a layer thickness of 27 nm at a distance $d$ of 3~nm. We also observe a high nitrogen incorporation, with a peak concentration of 208~ppm, resulting in an incorporation efficiency of $2.9\cdot10^{-3}$. These effects could be related to a lower substrate temperature in these lower positions, which generally favors nitrogen incorporation~\cite{lobaev2017influence, yan1999multiple, tallaire2006multiple}. Additionally, the lower position might be protected from direct exposure to the plasma, and thus allowing fewer reactive species to reach the sample, lowering the growth rate. This change in the growth regime is beneficial for the fabrication of delta-doped layers, because the low growth rate enables precise adjustment of the layer thickness. Furthermore, the high nitrogen concentration could enable high-performance sensors, due to the high NV concentrations~\cite{teraji2024nitrogen}.

As can be seen in Fig. \ref{fig_sims_overview}, the first layer, which was fabricated at a distance of 20~mm from the top level of the baseplate, shows that a significant amount of nitrogen was incorporated (blue curve), even though the sample was not in contact with the plasma. This suggests that the reactive plasma species are mobile and have a sufficient lifetime outside the visible plasma ball to reach the sample 2~cm below. We also observe a high nitrogen incorporation in these growth conditions, reaching a peak concentration of 12~ppm. Interestingly, no dip in the \textsuperscript{13}C concentration of this layer is observed, suggesting that no significant diamond growth occurs at this depth during the growth duration of 5~min.

The nitrogen in this layer might also have been incorporated during the subsequent intrinsic growth process. While nitrogen is in the plasma, it could be deposited on the diamond surface and terminate it.  In literature, it has been shown that the overgrowth of a nitrogen-terminated surface can lead to the incorporation of nitrogen in the growing diamond~\cite{jaffe2020novel, chandran2016fabrication}.

\begin{figure}[t]
\centering
\includegraphics[width=0.99\linewidth]{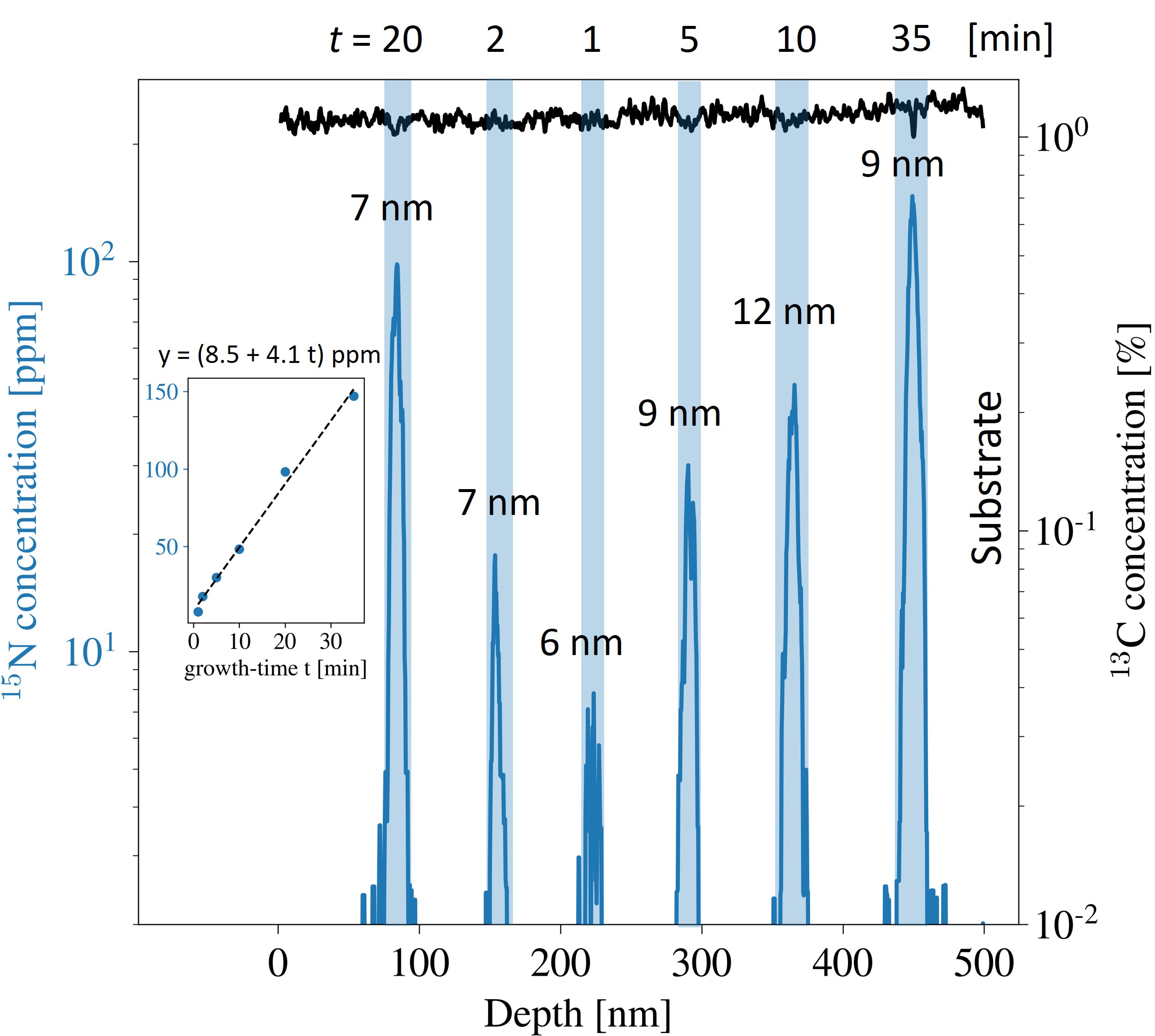}
\caption{ToF-SIMS profile of sample S2, where the growth time at a distance $d=20~$mm was varied (35, 10, 5, 1, 2, 20~min from substrate), as indicated for each layer above the graph. No dip in the \textsuperscript{13}C concentration is observed (black), which indicates a negligible diamond growth rate when the sample is in this position, with nitrogen as the source gas. The nitrogen incorporation in the layers (blue) increases linearly with increasing growth time, as shown in the inset.}
\label{fig_sims_overview2}
\end{figure}

To further investigate this behavior, we grew the same six-layer stack again, but kept the position at 20~mm distance and varied the process time from 1~min to 35~min. The resulting depth profile is shown in Fig. \ref{fig_sims_overview2}. There is no significant or linear increase of the layer thickness, indicated by the increase in nitrogen concentration, with respect to the growth time. Additionally, the nitrogen concentration increases with increasing growth time, hinting that there is no constant growth and defect incorporation, which would be independent of time, but rather a deposition of nitrogen species. Furthermore, again, no significant dip in the \textsuperscript{13}C concentration is observed.

This growth concept challenges the common belief that the sample must be in direct contact with or very close to the plasma to allow single crystal diamond growth~\cite{gicquel2001cvd, mankelevich2008new, mokuno2005synthesizing, muchnikov2010homoepitaxial}. In fact, we observe diamond growth, as indicated by the change in isotopic composition, up to 10~mm from the top level of the baseplate. 

The different growth behavior opens up new avenues for diamond epitaxy. They allow for changes in the growth rate and defect incorporation in doped layers, while maintaining constant plasma conditions. Furthermore, the deposition of defect species in positions that are more than 10~mm away from the base plate could enable new techniques for color center fabrication.

We also used this alternative growth mode to incorporate other species, such as phosphorus. There, we again observed a high concentration spike, which was one order of magnitude higher than the incorporation in the reference position, where the sample is in the plasma. The associated data can be found in the Supplementary Material.

To assess the suitability of the fabricated layers for quantum-technological applications, the PL intensity of single layers that were grown in the different positions was measured. The same growth conditions as before were used, but the N/C-fraction in the gas phase was reduced to 10,400~ppm, to reduce the nitrogen concentration in the thin films facilitating the measurement of quantum properties by decreasing bath-related decoherence. The positions with respect to the baseplate were 0.1~mm (S3), 3~mm (S4), and 20~mm (S5). Based on previous processes in the reference position, we estimate the P1 center concentration in sample S3 to be 2~ppm. 

\begin{figure}[t]
\includegraphics[width=0.999\linewidth]{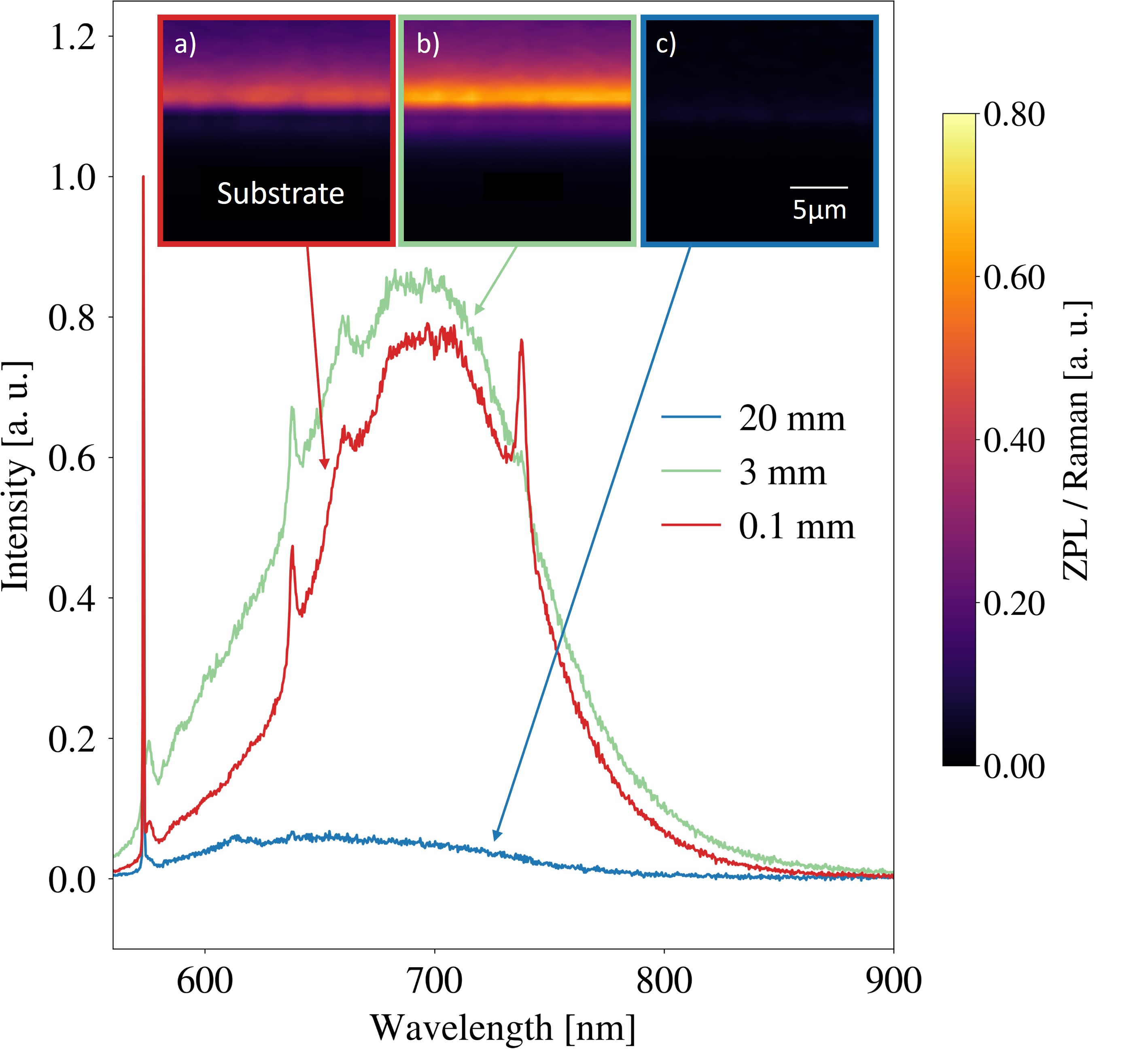}
\caption{Example spectra from bright spots of the PL slices, shown in a) to c). a) Depth slice through sample S3, grown 0.1 mm below the reference position. b) and c) Depth slice through sample S4 and S5, where the doped layer was grown 3~mm and 20~mm below the baseplate level. The color scale on the right side is valid for all plots, a) to c).}
\label{fig_PL}
\end{figure}

We then performed PL measurements, using a Renishaw InVia confocal microscope, as described in the Supplementary Material. We acquired depth slices within a 20~µm x 20~µm section of the sample, as shown in Fig. \ref{fig_PL} a) to c). The step size in the x- and z-directions was 1~µm and 0.5~µm, respectively. To compensate for fluctuations in laser power or optical losses, the signal acquired at the zero phonon line of the NV\textsuperscript{-} center (637~nm) was normalized with respect to the Raman peak at 573~nm.

The sample grown 0.1 mm below the reference position exhibits strong PL intensity and a clearly visible zero phonon line of the NV center. Additionally, SiV\textsuperscript{-} emission, with a zero phonon line at 737 nm~\cite{toth2015zero} was observed. Silicon is often incorporated during CVD diamond growth and could be introduced to the reactor via plasma etching of the glass bell jar~\cite{yang2024formation}. We assume that the high silicon incorporation arises from surface damage caused by etching processes during plasma changes, particularly when the sample is in close proximity to the reference position. These etching processes could lead to the formation of facets with enhanced defect incorporation~\cite{ashfold2020nitrogen}.

The sample, grown in a position 3~mm below, shows a similar PL intensity, despite the expected layer thickness being nearly seven times lower. This is most likely due to the high nitrogen incorporation, and confirms that this is an effective method for obtaining thin films with a high NV concentration. An example spectrum, as shown in Fig. \ref{fig_PL}, does not reveal the presence of a high concentration of SiV centers. However, the curve suggests the presence of both NV\textsuperscript{-} and NV\textsuperscript{0}, with a zero phonon line at 575~nm~\cite{manson2013assignment}.

The layer that was grown 20~mm below the reference position has the lowest nitrogen incorporation in the six-layer stack, and a thickness of only 19~nm. As expected, the PL intensity of sample S5 is lower than that of the other samples. The single graph in Fig. \ref{fig_PL} reveals a slight bump barely above the noise level, indicating the presence of only a few NV centers.

The coherence times of sample S4 were determined, using a home-built setup for measuring NV ensembles, as described in the Supplementary Material. We obtain T\textsubscript{2}* values of 0.38~µs and T\textsubscript{2} values of 7.1~µs. We assume, the values are mainly limited by the high nitrogen concentration. The measured T\textsubscript{2}-time implies a P1-center concentration of  23~ppm~\cite{bauch2020decoherence}. The layer grown at a nitrogen concentration in the gas phase roughly seven times higher exhibits a nitrogen concentration of 228~ppm, as determined by ToF-SIMS. This suggests a nitrogen concentration of 33~ppm in sample S4, if the incorporation efficiency was the same. The relatively higher coherence time could arise from a changing incorporation efficiency or the quasi two-dimensional nature of the layer~\cite{schatzle2024spin, bogdanov2021investigation}.

 In this work, we present a new diamond growth technique that allows the fabrication of nitrogen delta-doped layers with high nitrogen concentration. We demonstrate two distinct growth regimes that differ significantly from established processes. The photoluminescence measurements suggest a high NV concentration in the layers, fabricated in a retention position, 3~mm below the reactor baseplate, making them suitable for use in quantum sensing applications. Moreover, we observe reduced nitrogen incorporation during growth, 20~mm below the plasma, resulting in the formation of small ensembles of NVs. These layers, which have a low NV concentration could be beneficial in the field of quantum computing. Additionally, the new approaches may open up new avenues for diamond growth in various other areas. For instance, enhanced phosphorus incorporation could be advantageous for fabricating diamond-based electronic devices.  While the growth mechanism is not yet fully understood, we believe that ongoing investigations can elucidate this promising technique. We are currently studying the surface of the diamond after the sample has been positioned 20~mm below the reference position in order to gain a better understanding of the behavior and potential applications. The presented CVD processes could be scaled up to be applied to wafer-scale diamond samples with a higher diameter.

\section*{Supplementary Material}
The supplementary material includes a more detailed description of the growth parameters and the sample preparation. Additionally, results on phosphorus doping are shown.

\section*{Acknowledgements}
This work was supported by the european union, via the projects SPINUS (Grant No. 101135699) and AMADEUS (Grant No. 101080136). The authors thank Tristan Petit, Arsène Chemin and Thomas Dittrich for fruitful discussions.

\section*{Data Availability}
The data that supports the findings of this study are available from the corresponding author upon reasonable request.

%

\end{document}